\documentclass[epj,referee]{svjour}
\usepackage{graphicx}
\begin{document}
\title{Upgrade of the Glasgow photon tagging spectrometer for Mainz MAMI-C.}
\author{J.C. McGeorge\inst{1}\and J.D. Kellie\inst{1}\and
J.R.M. Annand\inst{1}\and J. Ahrens\inst{2}\and I. Anthony\inst{1}\and 
A. Clarkson\inst{1}\and D.J. Hamilton\inst{1}
\and P.S. Lumsden\inst{1}\and E.F. McNicoll\inst{1} 
\and R.O. Owens\inst{1}\and G. Rosner\inst{1}\and
A. Thomas\inst{2}
}                     
%
%

\institute{Department of Physics and Astronomy, University of Glasgow, Glasgow G12 8QQ,
Scotland, UK
\and Institut f\"{u}r Kernphysik, Universit\"{a}t Mainz, D-55099 Mainz, Germany
}
\date{Received: date / Revised version: date}
%
\abstract{
The Glasgow photon tagging spectrometer at Mainz has been upgraded so that
it can be used with the 1500 MeV electron beam now available from the
Mainz microtron MAMI-C. The changes made and the resulting properties of
the spectrometer are discussed.
\PACS{
      {29.40.Mc}{Scintillation detectors}   \and
      {29.30.Dn}{Electron spectroscopy}
     } 
} 
\titlerunning{\it{Tagger upgrade}}
\authorrunning{\it{J.C. McGeorge} et al.}
\maketitle
\section{INTRODUCTION}

\label{intro}

The Glasgow photon tagging spectrometer \cite{tag1,tag2}
installed at the MAMI-B 883~MeV electron microtron \cite{mami1,mami2,mami3}
at Mainz, Germany in 1991 has been used in many successful photonuclear
experiments. 
The main focal plane detector \cite{tag2} consisting of 353 plastic
scintillators covered a tagged photon energy range of
$\sim$40 - 820~MeV at full MAMI-B energy
and allowed a maximum
tagged photon flux of 
$\sim$5$\times10^5$ per MeV $\cdot$ s. 
Although the intrinsic resolution of the spectrometer was
$\sim$0.1~MeV \cite{tag1} the effective resolution was
$\sim$2~MeV due to the widths of the detectors.
Improved resolution
over part of the energy range was provided by a 96-element focal
plane microscope \cite{uscope}. Using an aligned diamond radiator
tagged photons with linear polarisation greater than 45\% have been
produced \cite{poln1,poln2,poln3,natter,poln4}
over an adjustable part of the energy range up to $\sim$400~MeV.
Circularly
polarised tagged photons were also generated using polarised electrons
from MAMI-B \cite{circpol}.
Several powerful detector systems such as Daphne \cite{daphne},
CATS \cite{cats}, PIP/TOF \cite{piptof}, TAPS \cite{taps} and most recently
the Crystal Ball \cite{cb} have been used in conjunction with MAMI-B and
the Glasgow tagger to make measurements on meson photoproduction and
to study photonuclear reactions. Examples include studies of the
Gerasimov-Drell-Hearn sum rule \cite{gdh}, the E2/M1 ratio in
the N $\rightarrow \Delta$ transition \cite{beck} and two
nucleon knockout with linearly polarised photons \cite{g2n}.

The recent upgrade of the MAMI accelerator to 1500 MeV, in principle, gives
access to interesting tagged photon experiments at higher energy.
The photon linear polarisation can also be improved to
$>$60\% up to $\sim$800~MeV by using the tighter collimation 
allowed by the smaller opening angles in the Bremsstrahlung process at
higher energy.
Examples of such experiments are the detailed study of the second resonance
region with complete measurements on pseudoscalar meson production,
rare $\eta$ decay modes to look for physics beyond the
standard model and strangeness production near threshold to test
chiral perturbation theory.
But, as the maximum attainable magnetic field in the original
spectrometer was not sufficient to handle an electron energy of 1500 MeV,
major modifications were necessary. This paper describes these
modifications and the resulting properties of the upgraded
spectrometer.

\section{AIMS OF THE UPGRADE}
\label{sec:aims}

The original spectrometer deflected the 833~MeV electron beam
through $\sim$79$^0$ into a beam dump recessed into
the wall of the experimental hall, and the
tagged photons passed into a large well shielded experimental area.
It was decided, as far as possible, to preserve the original layout 
and also the original spectrometer optics which govern the
excellent and well
understood performance. This in turn avoided the costly redesign
of the focal plane detector and its mounting frame. In
consequence the magnetic field in the spectrometer had to be
increased from 1.0 to $\sim$1.8~T in order to deflect the 1500 MeV
beam into the original beam dump.

The original focal plane scintillators and electronics
were, however, replaced since the light output had become reduced,
typically by a factor greater than 10, due to radiation damage and
the original electronics had become obsolete and
incompatible with the CATCH \cite{CATCH} electronics used with the
Crystal Ball detector system. 
As the detector geometry was not changed the effective resolution
becomes $\sim$4~MeV when tagging at 1500~MeV.

\section{UPGRADE OF THE SPECTROMETER MAGNET}
\label{sec:magnet}

\begin{figure*}[htb]
\begin{center}
\includegraphics[scale=0.5]{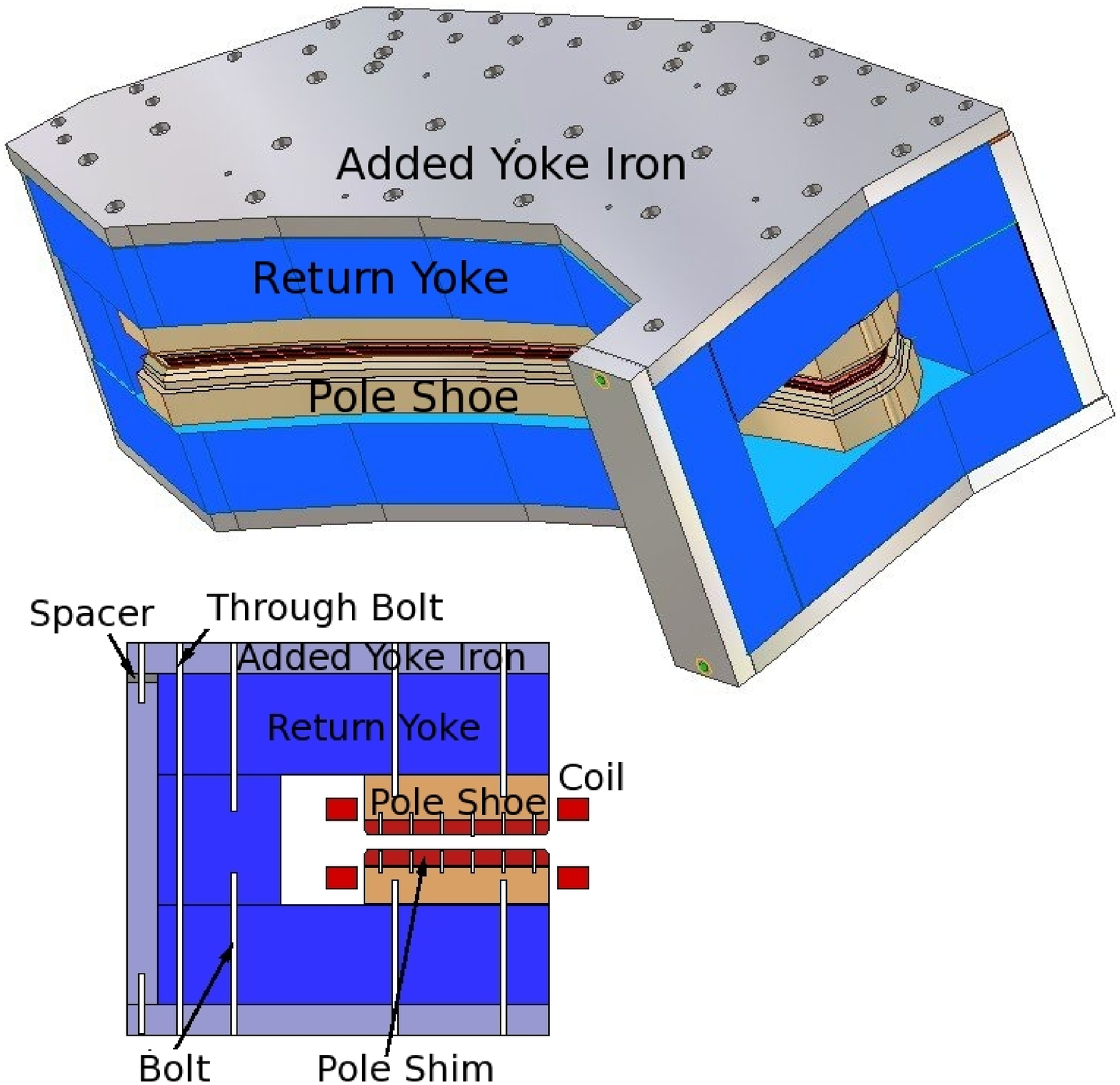}
\caption{
The upgraded photon tagging spectrometer - 3D view (upper) and cross section
(lower).}
\label{fig:ug1}
\end{center}
\end{figure*}

The existing power supply and cooling arrangements for the magnet coils
allowed for a current up to 440~amps which produced a field of
1.4~T \cite{tag1}. Simple estimates
suggested that 1.8 T could be obtained by reducing the pole
gap as long as the iron in the return yoke was increased in thickness to
prevent saturation.  
Reducing the pole gap is permissible since
angles associated with the Bremsstrahlung process scale approximately
as 1/energy, and a reduction from
50~mm (for~883 MeV) to 25~mm (for 1500~MeV) does not appreciably increase
the fraction of the
post-Bremsstrahlung electrons (tagging electrons) which hit the pole faces
of the magnet. With a 25~mm pole gap and 110~mm extra return yoke
thickness (Fig \ref{fig:ug1}),
calculations using the finite element code TOSCA \cite{tosca} showed
that an average field of $\sim$1.96~T would be reached with a current
of 440~amps. Therefore new additional
top, bottom and back yokes were fabricated, each of thickness 110~mm.

Measurements \cite{tag1} had shown that the maximum pole gap distortion
in the original spectrometer was
0.14~mm at a field of 1.0~T, and this could be expected
to increase to $\sim$0.54~mm at 1.96~T. Some estimates
of the stresses and distortions involved were made by
finite element modelling (FEM) of the original and upgraded
spectrometers as
solid structures using the IDEAS \cite{ideas} and ABAQUS \cite{abaq}
programs.
The upgraded spectrometer was strengthened by replacing the main load
carrying bolts with through-rods (Fig  \ref{fig:ug1}) fitted with
special nuts which allow a controlled pre-tension to be applied by
pneumatic means.

To reduce the pole gap, 12.5~mm thick shim plates (Fig. \ref{fig:shim}) were
fabricated. The input and all output edges were cut at
an angle of 36.1$^0$ to continue the chamfer of the existing
poles. The shims were fixed to the poles by 129 M8 screws.
 
The new magnet parts were all manufactured from low carbon steel
(S275). Although this has a somewhat inferior B/H curve
compared to the material (AME2SX1) used in the original spectrometer,
it is not very different in the region above 1.5~T.
B/H curves were measured (by the Woolfson Centre for Magnetics
Technology, Cardiff, UK) for samples of the material used to make the
new parts and input to TOSCA to estimate the magnetic field.

The vacuum box was modified to provide a larger aperture for the two
NMR probes required to cover the extended field range and also to
increase the acceptance of the spectrometer to $\sim$100~milliradians
in the horizontal plane.
This is useful for M\o ller polarimetry which can be used to measure the
electron beam polarisation.
 
In the upgraded spectrometer the maximum pole gap distortion at full
current was found to be
$\sim$0.42~mm. This is slightly less than predicted from scaling up
the distortion measured in the original spectrometer and suggests
that the through-rods had a beneficial effect.

The photon collimator was re-aligned on the input beam direction
with an accuracy of $\sim$0.2 mm
using the adjustments built into the V-shaped collimator mounting block.
The focal plane detector
support structure is fixed to the magnet, but it was
surveyed to check that it was remounted in the
same position as before to an accuracy of about 1~mm.

\section{MAGNETIC FIELD MEASUREMENTS}
\label{sec:field}

\begin{figure*}[htb]
\begin{center}
\vspace{9pt}
\includegraphics[angle=90,scale=1.0]{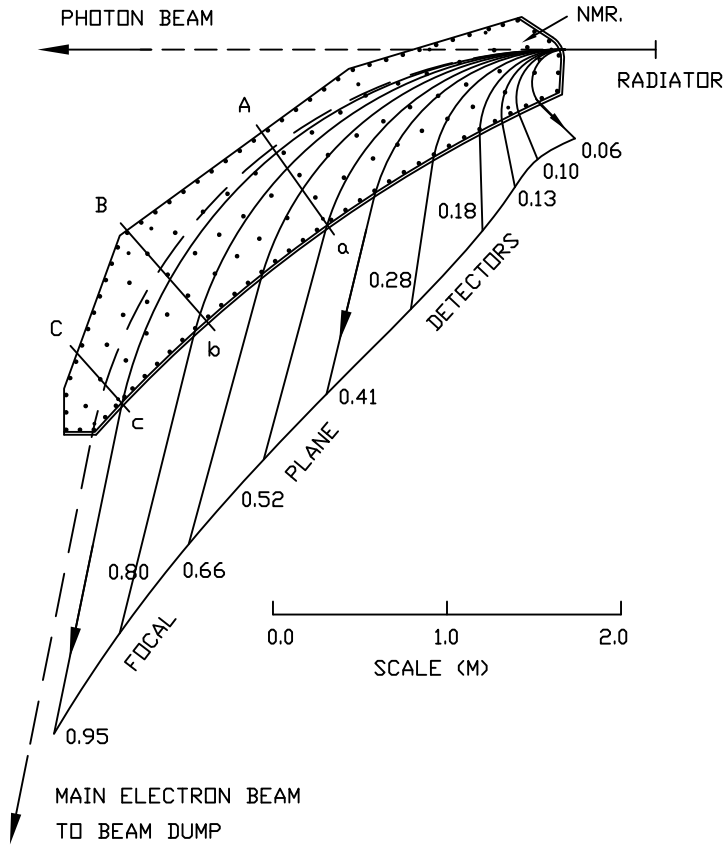}
\caption{Plan drawing of the lower pole shim (the upper pole shim is similar)
showing the locations of the M8 screws which fix it to the pole (dots).
The photon beam, main electron beam, several tagging electron trajectories
(labelled by their energy as a fraction of the main beam energy)
and the location of the main focal plane detectors are also indicated.}
\label{fig:shim}
\end{center}
\end{figure*}

\begin{figure*}[htb]
\begin{center}
\vspace{9pt}
\includegraphics[angle=-90,scale=0.5]{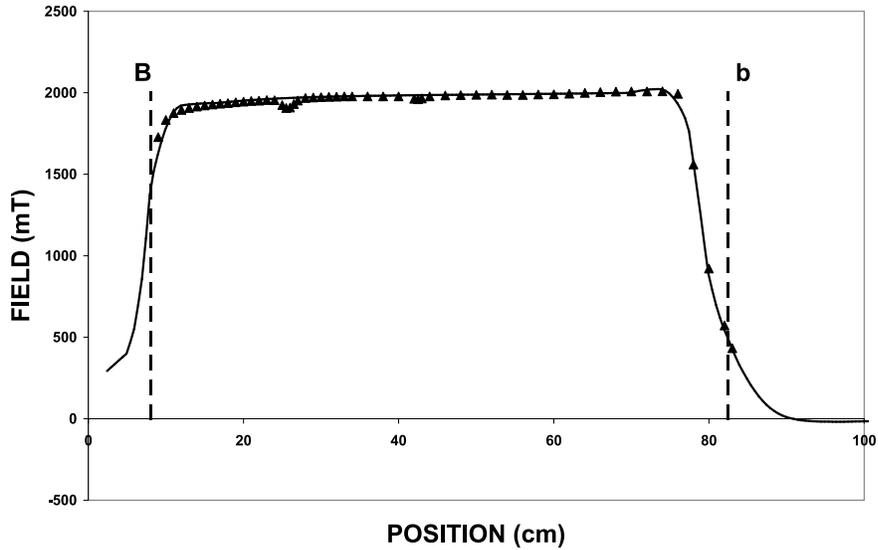}
\caption{The magnetic field at 435 A in the upgraded spectrometer
measured along line Bb in Fig. \ref{fig:shim} compared to the TOSCA
prediction (line). The dashed lines show the position of the pole edges.}
\label{fig:field}
\end{center}
\end{figure*}

Calculations of the magnetic field in the upgraded spectrometer made
with TOSCA indicated that the field would become more non-uniform as
the field is increased. As a result 
the energy of the tagging electrons, expressed as a fraction of the
incident beam energy, reaching a particular position in the focal plane
is expected to change slowly as the beam energy increases.
The spectrometer resolution should be less affected since the opening
angle of the tagging electrons is small.
Since  both the energy calibration and the energy resolution of the
spectrometer can be measured using electron beams of accurately known
energy from MAMI, complete field maps of the upgraded spectrometer at
several fields were not made. However some field measurements were made
using  a temperature 
compensated Hall probe which had been calibrated against an NMR system.
The field measured at mid-pole gap along line Bb in
Fig. \ref{fig:shim} is shown in Fig. \ref{fig:field}. It can be seen that the
measured field at 435~amps comfortably exceeds the 1.8~T required to handle
1500 MeV. It is in good agreement with the TOSCA prediction,
both at the edge from which the electrons exit and in the central
region, except
for the 'dips' at positions 25.5 and 42.5~cm corresponding to the locations
of two of the M8 screws which secure the pole shims.
Similar results were found along lines Aa and Cc.

It is not clear whether the 'dips' are predominantly due to
saturation of the screw material or to the missing metal which
arises from the clearance at the end of the tapped holes. Measurements
over a few screws indicate that the field reduction has a peak value
of $\sim$3.0\% and a full width at half maximum of about 17.5~mm.
Again, the effect on the energy resolution is expected to be  small
but there will be local deviations from a smooth energy
calibration. This is discussed in Sect. \ref{sec:tcal}.

It was not possible to re-optimise the field clamps for the reduced
pole gap without large scale mechanical reconstruction, and
the effect on the field edge was therefore examined carefully.
Both the field measurements and the TOSCA calculations indicate 
(Fig. \ref{fig:field})
that in the upgraded spectrometer the field edge is
displaced inward from the physical pole edge (see Fig.4 in ref \cite{tag1}).
This was quantified by determining the effective field boundary (EFB)
position for five locations based on the field calculated using TOSCA. The
EFB position was found by making the field integrals, $\int$B.dl, equal for
the uniform
and calculated fields along lines perpendicular to the pole edge, where the
magnitude of the uniform field is taken to be the average field
calculated for
the region 100-200~mm inside the physical pole edge. The EFB's were
found to be 21 $\pm$1~mm inside the physical boundaries
along lines Aa, Bb and Cc in Fig. \ref{fig:shim}.

\section{FOCAL PLANE DETECTOR}
\label{sec:fpd}

\subsection{The detectors}
\label{sec:fpd1}

The focal plane (FP) of the tagger dipole magnet is instrumented
(see Fig. \ref{fig:shim} and Figs. 1 and 3 in ref \cite{tag2}) with
353 overlapping plastic scintillators which cover an energy
range of around 5 - 93\% of E$_{0}$, the energy of the primary
electron beam. The scintillators are mounted in milled slots
to define their positions and angles with respect to the tagging
electrons, which are momentum analysed by the dipole magnet.
The scintillators
have a length of 80~mm, a thickness of 2~mm and widths of
9 to 32~mm. These decrease along the focal plane in order to keep
the tagged energy range covered by each detector roughly constant.
The scintillator strips overlap by slightly more than half
their width (see Fig. 3 in ref \cite{tag2})
so that an electron hit is defined by coincident signals
in adjacent detectors. The width of the overlap region (a 'channel')
is equivalent
to an energy width of $\sim$4~MeV, for an incident electron beam
energy of 1500~MeV, and neighbouring channels overlap by about 0.4~MeV.

The scintillator EJ200 was chosen for the refurbishment because the
scintillation spectrum
better matches the response of the phototube and it is thought to be less
susceptible to radiation damage than the slightly faster NE 111/Pilot U
used in the original setup.
The 353 new scintillators were glued, using ultra-violet curable epoxy,
to new light guides made from acrylic which has good transmission at blue to
near ultra-violet wavelengths. They were
then wrapped in double-sided, aluminised Mylar to eliminate
optical cross talk and mounted in the original detector
frame \cite{tag2}.

Before installation about half of the scintillators were tested
using a Sr90 source
and the signal amplitude was found to decrease linearly with increasing width.
Compared to the old detectors, the new scintillators produced one to
two orders of magnitude more light, due mainly to radiation damage accumulated
by the former in around 15 years of service.

\subsection{Photomultiplier amplifier-discriminator electronics}
\label{sub:electronics}
Each scintillator on the FP is fitted with a Hamamatsu R1635
photomultiplier tube (PMT). As the PMTs are affected by stray fields of
more than $\sim$0.01~T, 0.2~mm thick mild steel plates were fitted
along the whole length of the detector
array on either side of the PMTs. Together with the standard
cylindrical $\mu$-metal screens fitted to every PMT, this was found
to be sufficient to cope with the increased stray field from the
upgraded magnet when operated at maximum field.

Most of the original PMTs are still in
good working order. A total stock of around 450 was sorted on the basis
of gain, with the highest gain tubes fitted to the low-momentum end of
the spectrometer, grading progressively down in gain along the length of
the focal plane. This partially compensates for the less efficient light
collection from the relatively broad scintillators at the high-momentum end.

\begin{figure*}[htb]
\begin{center}
\includegraphics[scale=0.6]{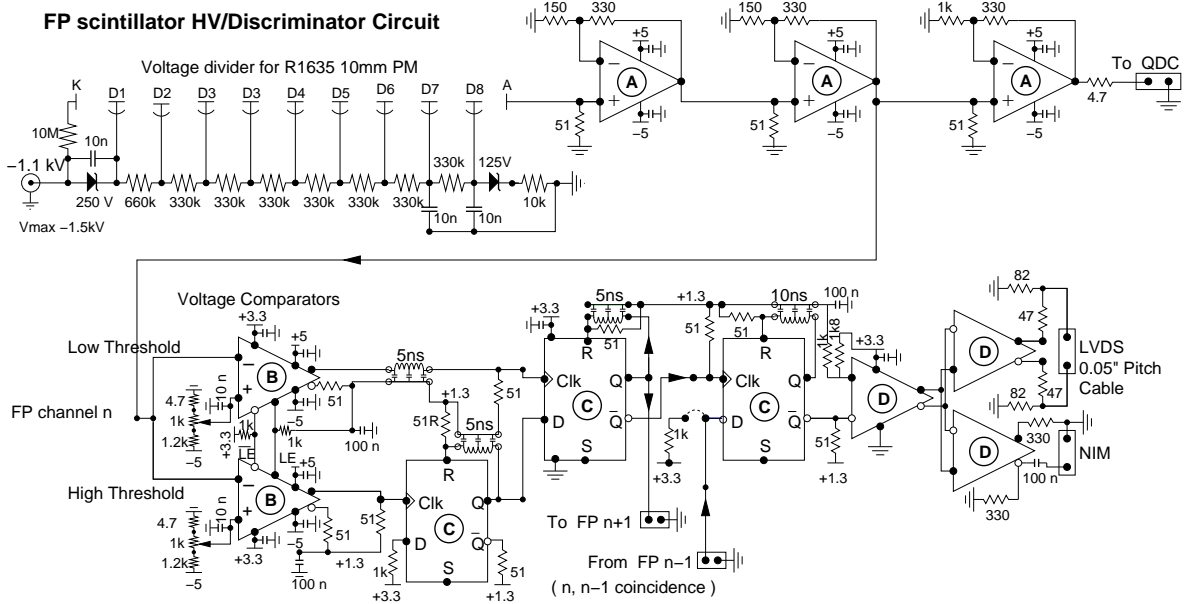}
\caption{Circuit diagram of the amplifier-discriminator card. Integrated
circuits are labelled A: AD-8009 1 GHz, current-feedback
operational amplifier; B: MAX-9601 dual ultrafast comparator; C: MC100LEVL30
triple D-type flip flop with S/R;  D: MC100LVEL11  buffer, fan out.
}
\label{fig:A/D-card}
\end{center}
\end{figure*}

Every PMT is attached to a custom designed amplifier-discriminator
(A/D) card
(Fig.\ref{fig:A/D-card}). High voltage (HV) is distributed to the
PMT electrodes through a Zener stabilised base chain which may be
operated from $900-1500$~V.
Typically a PMT is run at around -1100~V, drawing a current of around
0.3~mA, while each A/D card draws $\sim$370~mA from
the +5~V and $\sim$250~mA from the -5~V LV supply lines.

The anode signal is amplified by a factor 10 and fed to a dual, low-high
threshold discriminator which supplies a LVDS-logic signal to drive
TDCs and scalers.
The differential LVDS signal is transported on $\sim$10~m
of 0.05''-pitch cable, to active fanout cards which
connect to sampling, multi-hit TDCs and scalers (designated CATCH).
These were originally designed \cite{CATCH} for the COMPASS experiment
at CERN. The sampling TDCs, which have a channel width of 0.117 ns and
double pulse resolution of 20 ns, remove the need for delay in the TDC
input lines, which was necessary in the original setup
to accommodate the delay in the trigger system. The fine-pitch cable, which is
commonly used to connect SCSI-bus peripherals, is much
less bulky than standard 0.1'' pitch cable and, over a 10 m length, produces no
significant degradation of edge speed in the 10 ns wide pulse. In the previous
implementation of the tagger electronics, pulse widths had to be at least
20 ns to drive the $\sim$100 m of delay cable before the TDC input.
An active logic fanout is necessary not only to drive CATCH modules,
which cannot be ``daisy-chained'', but also for auxiliary ECL-logic
electronics, such as used to select M\o ller-scattering events for
electron-beam polarisation analysis.

\begin{figure*}[htb]
\begin{center}
\includegraphics[scale=0.25]{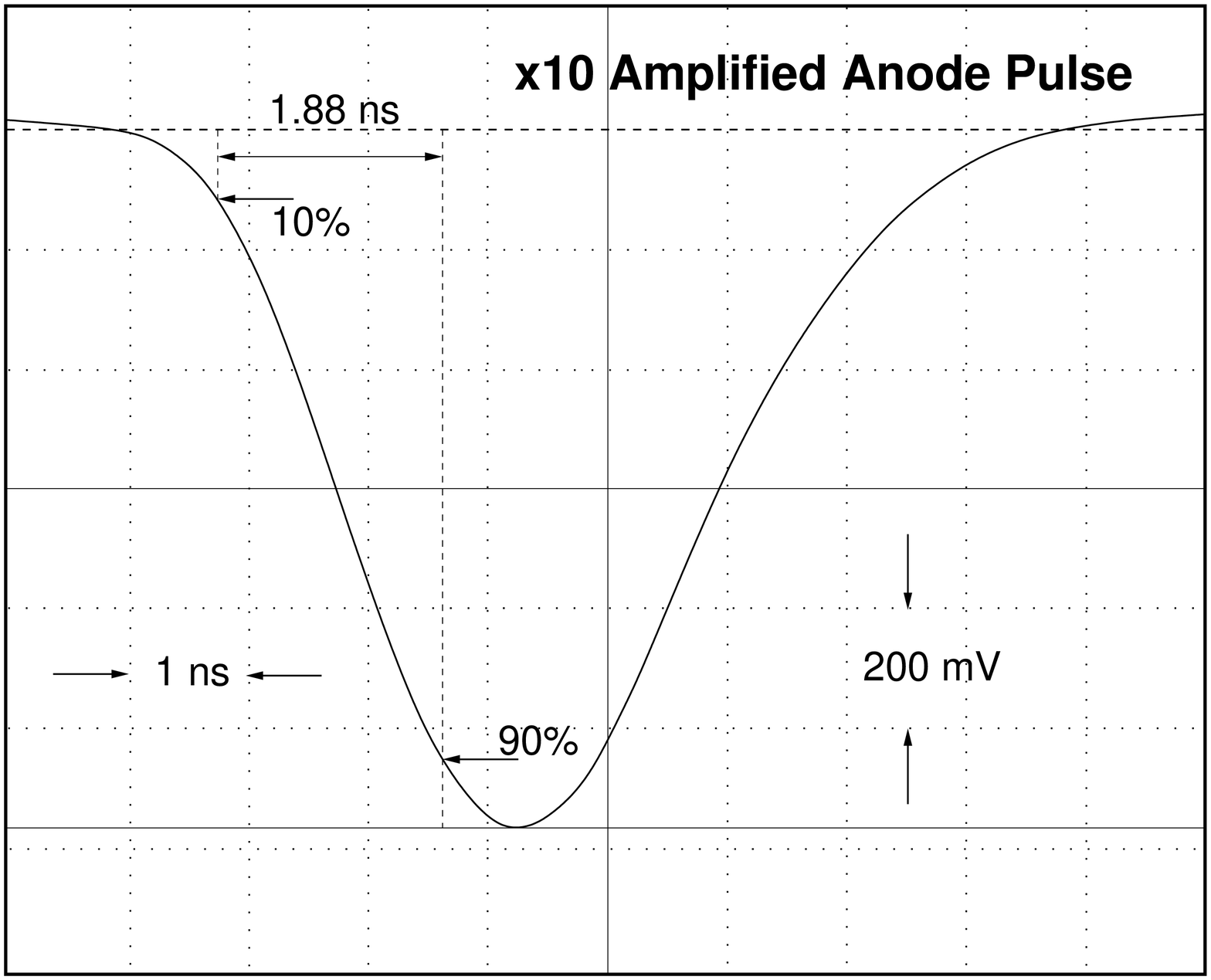}
\caption{Amplifier output.}
\label{fig:Amplifier-output}
\end{center}
\end{figure*}

The $\times10$ amplified anode signal,
produced by illumination of the R1635 PMT photocathode by a 1~ns
duration diode laser, is displayed in Fig. \ref{fig:Amplifier-output}.
The amplifier produces no discernible degradation of edge speed and
the 10-90\% rise time is 1.9~ns.
This output also connects, via a $\times1.3$ buffer stage and a coaxial
delay line, to a LeCroy 1885F FASTBUS QDC which is normally read out
during HV adjustment to align FP detector gains.

\subsection{\label{sub:Detector-performance}Detector performance}

\begin{figure*}[htb]
\begin{center}
\includegraphics[scale=0.6]{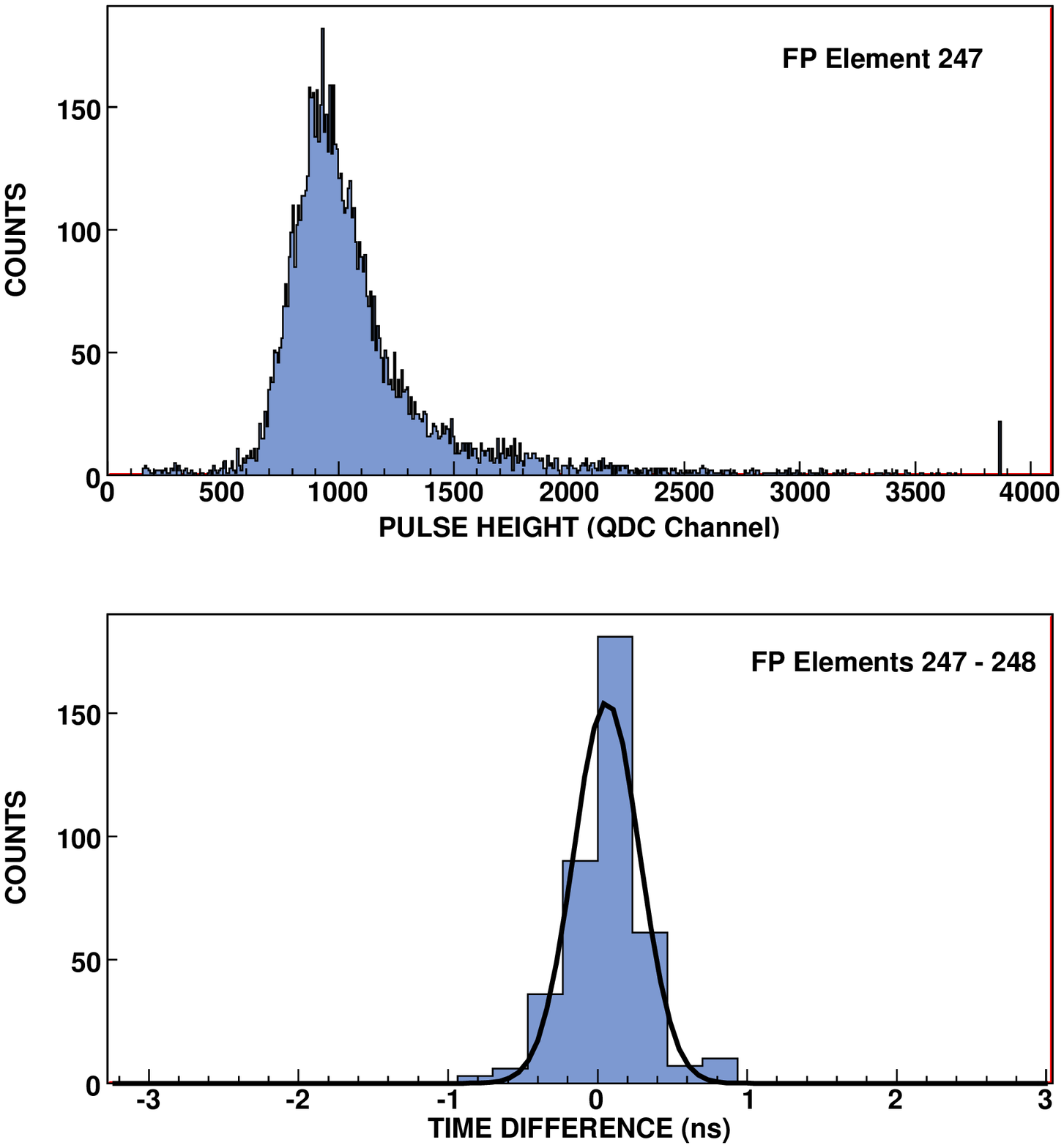}
\caption{Focal plane detector performance.}
\label{fig:FP-detector-performance}
\end{center}
\end{figure*}

The performance of the FP detectors and electronics is illustrated
in Fig. \ref{fig:FP-detector-performance} which displays spectra
taken with the MAMI-C 1508 MeV electron beam incident on a 10~micron thick
Cu radiator. Bremsstrahlung photons, detected by a lead-glass Cherenkov
detector placed directly in the beam,
triggered the data acquisition system and provided gates for the
charge integrating QDCs and the time reference for the TDCs.

The upper plot shows the pulse height spectrum produced by the tagging
electrons in a single FP detector. This has very
little background below the Landau distribution showing that detector noise
levels are well below the minimum ionising signal. It also suggests that
any uncharged background, for example from electron-beam interactions
with beam-line components is suppressed very effectively.

The lower plot of Fig. \ref{fig:FP-detector-performance} shows the
difference in hit times for two adjacent FP channels when an electron
passes through the small region where the channels overlap. The time
variation of the lead-glass trigger cancels in this difference so that
the observed width of the distribution for
elements $i$ and $i+1$ is $\delta t\simeq$
$\sqrt{\delta t_{i}^{2}+\delta t_{i+1}^{2}}$, where $\delta t_{i}$
is the timing uncertainty for element $i$.
Assuming $\delta t_{i}\simeq\delta t_{i+1}$,
the Gaussian-fit width ($\sigma)$ of 0.24~ns is equivalent to a single-counter
resolution of 0.17~ns (0.40~ns FWHM).
This performance is fairly typical, with measured single-counter widths
in the range 0.37 - 0.53~ns (FWHM) and is significantly better than
the pre-upgrade system where the best performance obtained
was $\sim$1ns FWHM.

\section{TAGGER ENERGY CALIBRATION}
\label{sec:tcal}

In the analysis of an experiment with the tagger it is necessary
to know the energy of the tagging electrons which hit the centre of each
focal plane detector channel.
As the MAMI beam energy can be measured with an uncertainty of
140 keV \cite{mami4} the tagger calibration can be carried out directly
using very low current MAMI beams of lower energy than in the
main experiment. However, it is only practical to obtain a small
number of calibration points in this way.

\begin{figure*}[htb]
\begin{center}
\includegraphics[scale=0.7]{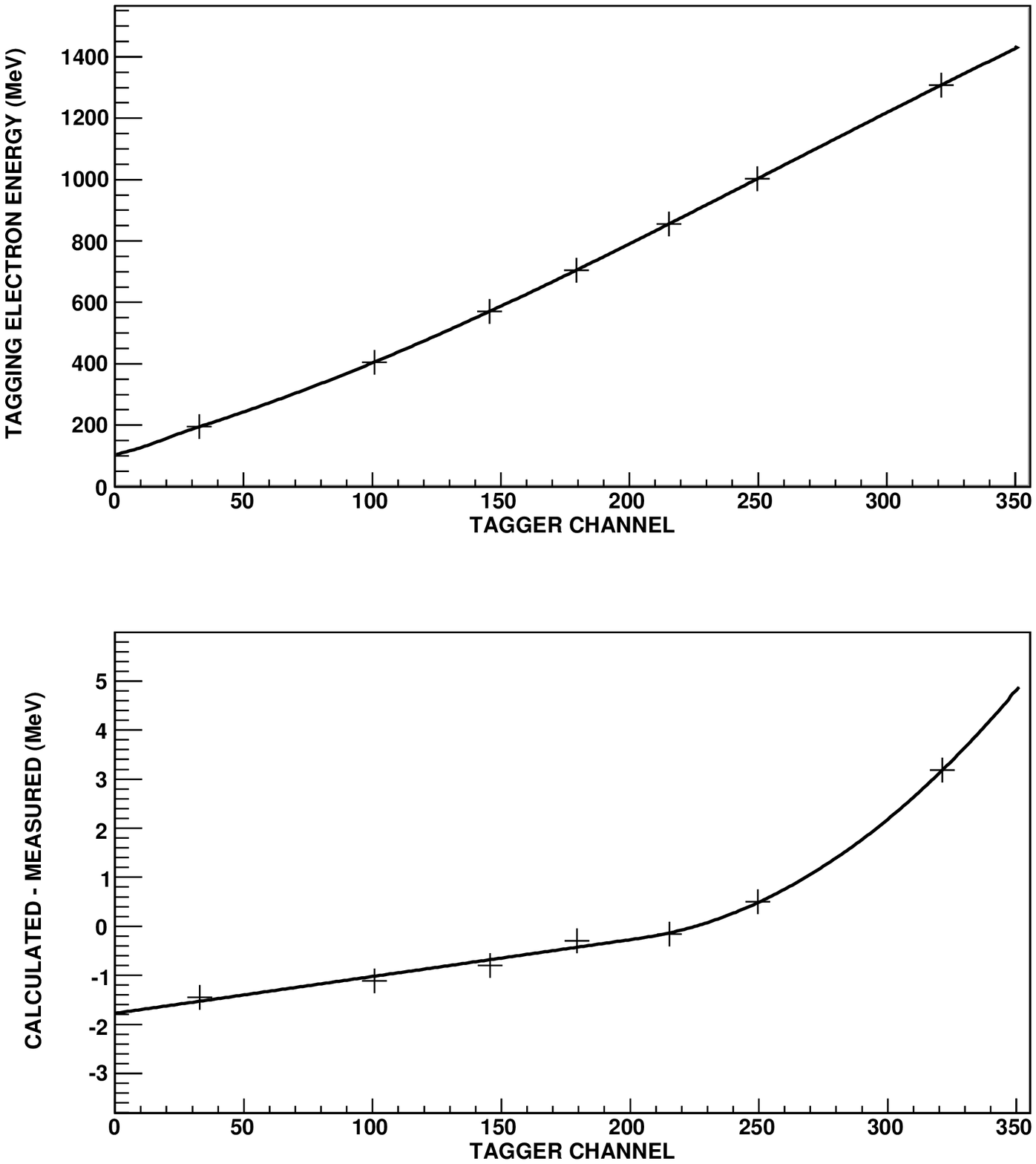}
\caption{Upper part: Tagger energy calibration for main beam energy
1508~MeV measured using MAMI energies 195.2, 405.3, 570.3, 705.3, 855.3,
1002.3 and 1307.8~MeV.
The line shows the calibration calculated assuming a uniform field.
\newline
Lower part: Difference between the calculated and measured calibrations.
The line here shows a smooth fit to the seven measured points and indicates
the small correction to the calculated calibration required because of
large-scale field non-uniformity.
}
\label{fig:tcal}
\end{center}
\end{figure*}

In the measurement the MAMI beam is 'scanned' across several 
(typically 12) focal plane
detectors by varying the tagger field slightly (up to $\pm$5$\%$) around the
value required to dump the beam correctly in normal tagging
(eg. around 1.834~T which is used when tagging with 1508~MeV).
By making fine steps it is
possible to measure the field values for which the beam hits the small
overlap regions  (see Sect. \ref{sec:fpd1}) between neighbouring channels.
This gives the hit position to an accuracy of about $\pm$0.05 channel. 
Interpolation of channel number versus field then gives the (fractional)
channel number hit for the correct field (1.834~T in the above example).
Such calibration measurements have been made with MAMI energies
195.2, 405.3, 570.3, 705.3 and 855.3, for a field of 1.057~T
(which is used to dump 833 MeV in normal tagging) and for
a field of 1.834~T where MAMI energies of 1002.3 and 1307.8 MeV
were also used. The resulting calibration for 1.834~T 
is shown in the upper part of Fig. \ref{fig:tcal} and
compared to that calculated assuming a uniform field as described below.
The difference between the measured and calculated calibration is shown
in the lower part of the figure.

If the field
shape along any electron trajectory were independent of field magnitude,
it would be possible to 'simulate' intermediate tagging electron energies,
E'= EB/B'  by varying the tagger field (B') away from the field B
for which the spectrometer is being calibrated using MAMI energy E.
The error that arises because the field shape is not
independent of field magnitude was investigated by making
several overlapping scans using different beam energies.
It was found that the error is too large for this method to provide
useful extra calibration information unless a suitable correction can
be applied.
While the assumption that the required correction is a linear function
of B-B' is thought to be sufficiently accurate over the small energy range
required for calibration of the microscope
detector (see ref \cite{uscope}), it may not be reliable over the wider
range required to provide extra, widely spaced, calibration points.
Therefore the calibrations were
based only on the seven (or five) points measured with the seven (or five)
different energies from MAMI at the correct
field of 1.834 T (or 1.057 T). To guide the interpolation between these points 
a computer program has been written to 
calculate the calibration on the basis of a uniform tagger field
with the effective field boundary determined in Sect. \ref{sec:field}.
The relative positions and angles of the scintillators
are known from the construction of the support frame \cite{tag2}
and its position relative to the magnet was determined by surveying. 
For electron trajectories made up of circular arcs and
straight lines the required calibration can be calculated by simple
geometry.
The strength of the field is taken from the 
value measured using a Drusche NMR system multiplied by a factor, f,
which accounts for the difference between the field at the NMR probe
(see Fig. \ref{fig:shim}) and the average field encountered by the
tagging electrons. The value
of f was adjusted to fit the measured calibration points.

For 1508 MeV the calculated calibration is within about 1.5~MeV of
the measurements over most of the energy range (lower part of 
Fig. \ref{fig:tcal}) but the discrepancy
increases to about 4~MeV for the lowest photon energies. This behaviour
is thought to be due to the large-scale non-uniformity of the field that can 
be seen partly in Fig. \ref{fig:field}. As the effect varies smoothly over the
tagged energy range the required correction to the uniform field model
prediction can be obtained by fitting a smooth line to the seven
points in the lower part of Fig. \ref{fig:tcal}.

Similar results have also been obtained for a field of 1.057~T. The
value of f was found to be 1.0098 for 1.057 T and 1.0003 for 1.834 T.

\begin{figure*}[htb]
\begin{center}
\vspace{9pt}
\includegraphics[angle=-90,scale=0.6]{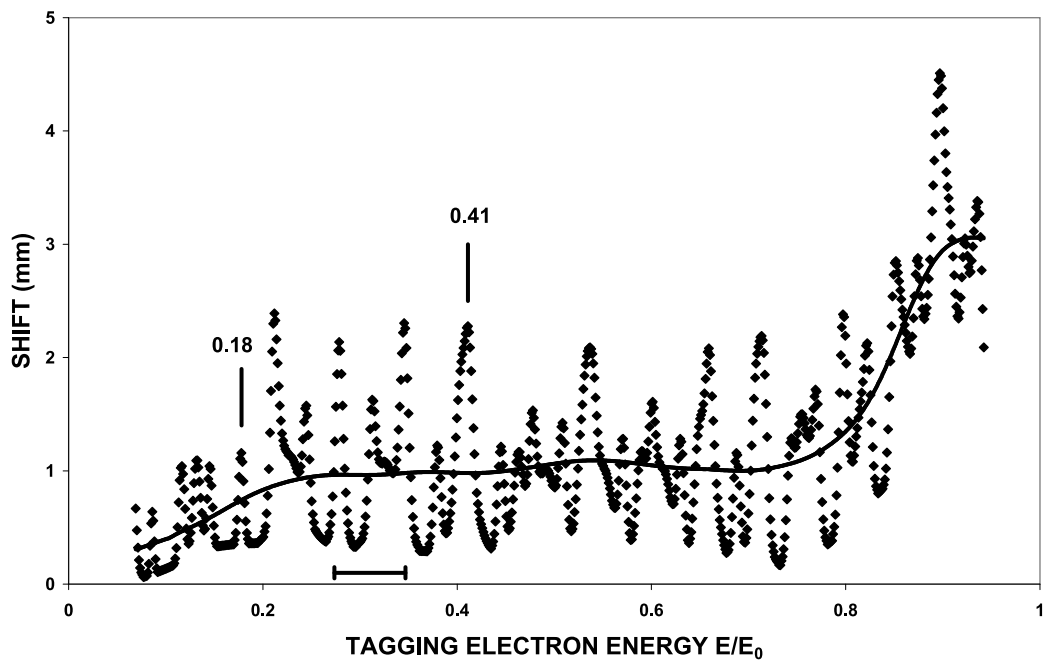}
\caption{Calculated shift of the tagging electron trajectories along
the focal plane due to the field dips caused by the M8 screws fixing
the pole shims. The line is the result of smoothing the points.
The horizontal bar shows the region covered by the
microscope for the energy calibration data shown in
Figs. \ref{fig:ucal} and \ref{fig:devn}.}
\label{fig:bobshft}
\end{center}
\end{figure*}

The deviations from a smooth calibration caused by the field dips due
to the pole shim mounting screws (see Sect. \ref{sec:field}) have been
investigated.
Estimated shifts (along the focal plane) of the tagging electron trajectories
brought about by the field dips are shown in Fig. \ref{fig:bobshft}
as a function of the tagging electron energy, E, expressed as a fraction
of the main beam energy, E$_{0}$.
This estimate was obtained by assuming the field is uniform between the
poles (except for the 'dips') and zero elsewhere and then
calculating the total effect on exit position and bend angle
caused by all dips whose centre is within 30~mm of the tagging
electron trajectory. It is assumed that the fractional effect
on bend angle is simply the fractional deficit in the field integral
compared to that for a uniform field with no dips.
The peaks in Fig. \ref{fig:bobshft} occur when the electron trajectory
passes over or near the centre of one or more M8 screws. For example,
as can be seen in Fig. \ref{fig:shim}, one screw near the output edge
lies near the E/E$_0$ = 0.18 trajectory,
and 3 screws lie on or close to the  E/E$_0$ = 0.41 trajectory.

The line in Fig. \ref{fig:bobshft} shows the result of smoothing the
shifts and values at the peaks and valleys are typically $\pm$0.6~mm different
from this line.
Although the trajectory shifts due to the 'dips' in the real field
may be different in detail from this simple estimate, 
Fig. \ref{fig:bobshft} implies that
the error that results from using a calibration
method where the calibration is assumed to be smooth
is about $\pm$0.2~MeV when tagging with a main beam
energy of 1500~MeV.
This is small compared to the 4~MeV channel
width of the main focal plane detector. However it
may be significant in experiments which use the focal plane
microscope \cite{uscope} where the channel width is $\sim$2~mm
along the focal plane. In such experiments a detailed energy
calibration can be performed by using one or two
different electron beam energies from MAMI and 'scanning'
them across the microscope by making small variations in the
magnetic field in the spectrometer (see ref \cite{uscope}).

\begin{figure*}[htb]
\begin{center}
\vspace{9pt}
\includegraphics[angle=-90,scale=0.5]{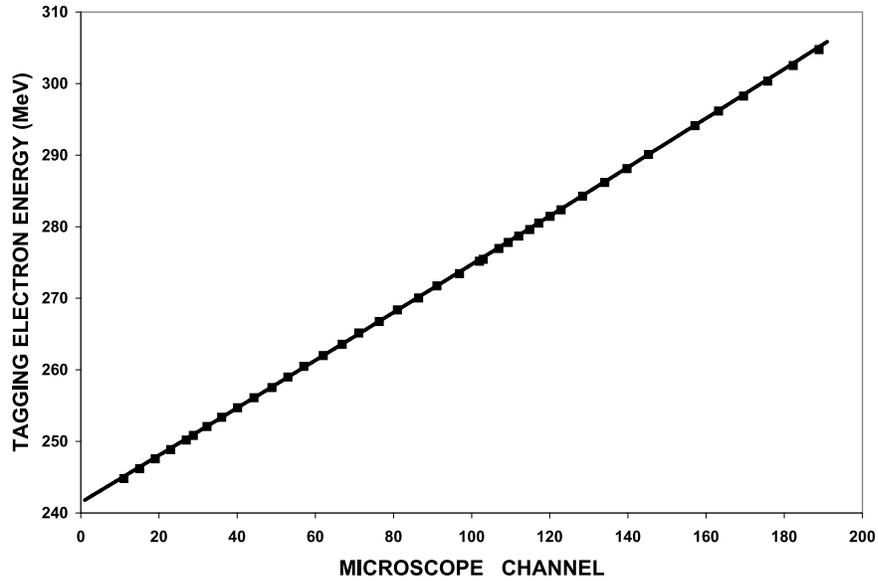}
\caption{Energy calibration in the region E/E$_0$ = 0.27 - 0.35 (for
E$_0$ = 883~MeV) obtained from scanning a 270.17~MeV beam from MAMI
across the microscope by varying the tagger field. The
line shows the result of a calculation assuming a uniform field.}
\label{fig:ucal}
\end{center}
\end{figure*}

\begin{figure*}[htb]
\begin{center}
\vspace{9pt}
\includegraphics[angle=-90,scale=0.5]{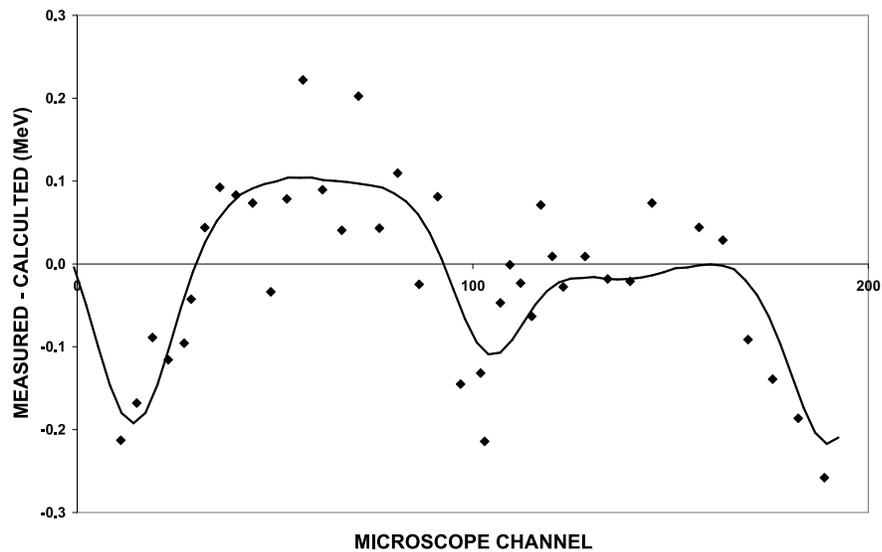}
\caption{Difference between the measured and calculated microscope
calibrations shown in  Fig. \ref{fig:ucal}. The line shows the calculated
difference (see text) based on the appropriate section of
Fig. \ref{fig:bobshft}.}
\label{fig:devn}
\end{center}
\end{figure*}

Such a scan can also be used to look for the effects of the shifts predicted
in Fig. \ref{fig:bobshft} over the small range covered by the
microscope detector.
This has been done with the microscope covering the range
E/E$_{0}$ = 0.27 - 0.35 (indicated by the horizontal bar in 
Fig. \ref{fig:bobshft}).
In Fig. \ref{fig:ucal} the measured microscope calibration points are compared
to the calibration calculated assuming a uniform field and using
the known microscope geometry.
As the microscope position and angle
were not known with sufficient precision these were adjusted in
the calculation to fit the measured points. The difference between
the measured points and the calculation is shown in detail in
Fig. \ref{fig:devn}. The line in this figure shows the
difference between the original points and smoothed line in the
relevant section of Fig \ref{fig:bobshft}.
The agreement in Fig \ref{fig:devn} is good enough to give some confidence in
the estimate, made above, of the energy calibration error arising from the
assumption that the calibration is smooth.

Including the uncertainty in the MAMI beam energy the uncertainty in
the seven calibration points (measured for tagging at a main beam
energy of 1508~MeV) is estimated to be about $\pm$0.3~MeV.
Measurements of the pole shim thicknesses suggest that
small variations in the pole gap could cause slight structure in the
calibration between the measured points. Including this, it is
estimated that the error in the calculated calibration, after
correction for large-scale field non-uniformity using the fit
shown in the lower part of Fig. \ref{fig:tcal}, is about $\pm$0.5~MeV
for channels up to $\sim$270. It could be significantly larger for lower
tagged photon energies where the shape of the correction is less well
defined. There is also an additional uncertainty caused by the
field dips. From Fig \ref{fig:bobshft}, this is estimated to be
typically about $\pm$0.2~MeV.

\section{THE PERFORMANCE OF THE UPGRADED SPECTROMETER}
\label{sec:tests} 

The intrinsic resolution of the upgraded
spectrometer was measured at a field of 1.95~T which made an 855 MeV beam
from MAMI hit the microscope detector placed near the middle of the focal plane.
Multiple scattering in a 2~mm thick Al sheet  placed in the beam at the
radiator position was used to simulate the opening angle distribution of tagging
electrons.
From the distribution of electrons hitting a small number of microscope
channels the
resolution was found to be $\sim$0.4 MeV FWHM. This is an overestimate because
the opening angles in this test were about 3 times bigger than is the case for
the tagging electrons from a typical radiator at 1508~MeV main beam energy.
It shows that when the
main focal plane detector is used its channel width ($\sim$4~MeV) dominates
the tagged energy resolution.
For some experiments which make use of the focal plane microscope, however,
it may be necessary to make more careful measurements of the resolution.

\begin{figure*}[htb]
\begin{center}
\includegraphics[angle=-90,scale=0.6]{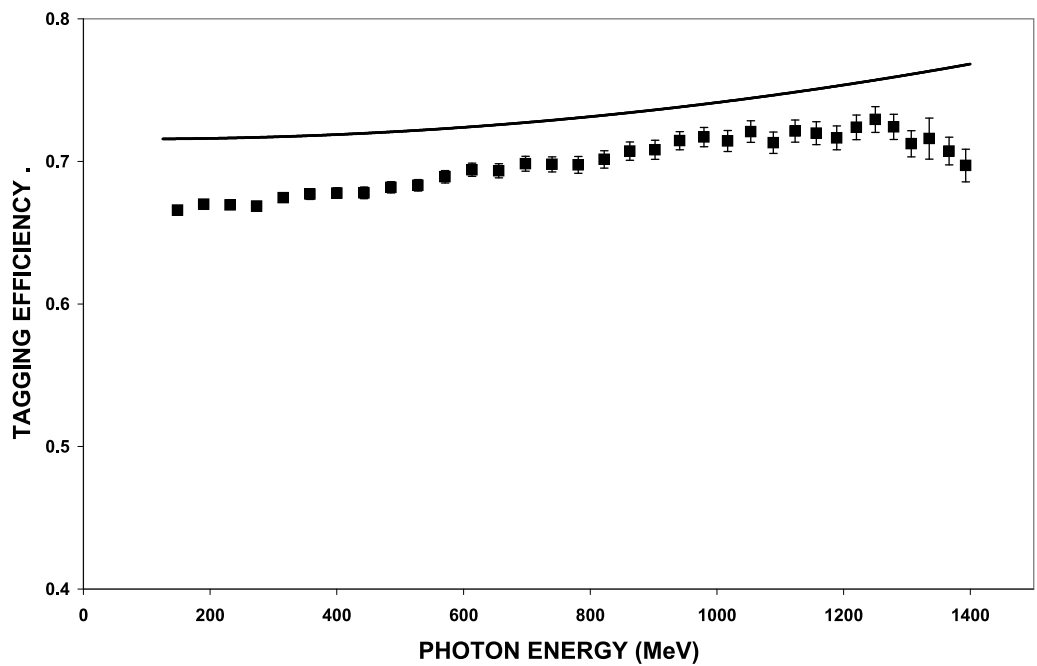}
\caption{Tagging efficiency measured at main beam energy of 1508~MeV using
a 10~micron thick Cu radiator and a 4~mm diameter collimator. The line
shows the result of a Monte Carlo calculation.}
\label{fig:etag}
\end{center}
\end{figure*}

Measurements of the 'tagging efficiency', that is the fraction of the tagged
photons which pass through the collimator, were made at reduced beam current
using a 25~cm$^3$ lead-glass Cherenkov detector placed on the photon
beam line. The results from a measurement using a 10~micron thick Cu radiator
and a 4 mm diameter collimator are shown in Fig. \ref{fig:etag}.
The measured tagging efficiency was found to be significantly smaller
than predicted
by a Monte Carlo calculation which includes the input beam
divergence and diameter, multiple scattering in the radiator,
the Bremsstrahlung photon opening angle distribution and the effect
of M\o ller electron scattering. A 
similar discrepancy was also found using a 
collimator diameter of 3~mm and a 6~micron thick Ta radiator.
Although not fully understood, much of the discrepancy may be due
to slight misalignment of the collimator.
The measured tagging efficiency is, however, stable and reproducible.
Measurements made several months apart averaged over all tagger channels
agree to better than 1\%.

As a test of the focal plane detector background,
the count rate was measured with no radiator in the beam. It 
was found to be about 5$\times10^{-5}$ times smaller than the rate
with a 10~micron thick Cu radiator in the beam.

The maximum useful tagged-photon intensity depends on the maximum rate
at which the FP counters can be run. Up to a rate of $\sim$1 MHz per channel
(detector singles rate of $\sim$2 MHz) no major change of pulse height
was observed, so that the detection efficiency for minimum ionising
particles did not change significantly.

\begin{figure*}[htb]
\begin{center}
\includegraphics[scale=0.6]{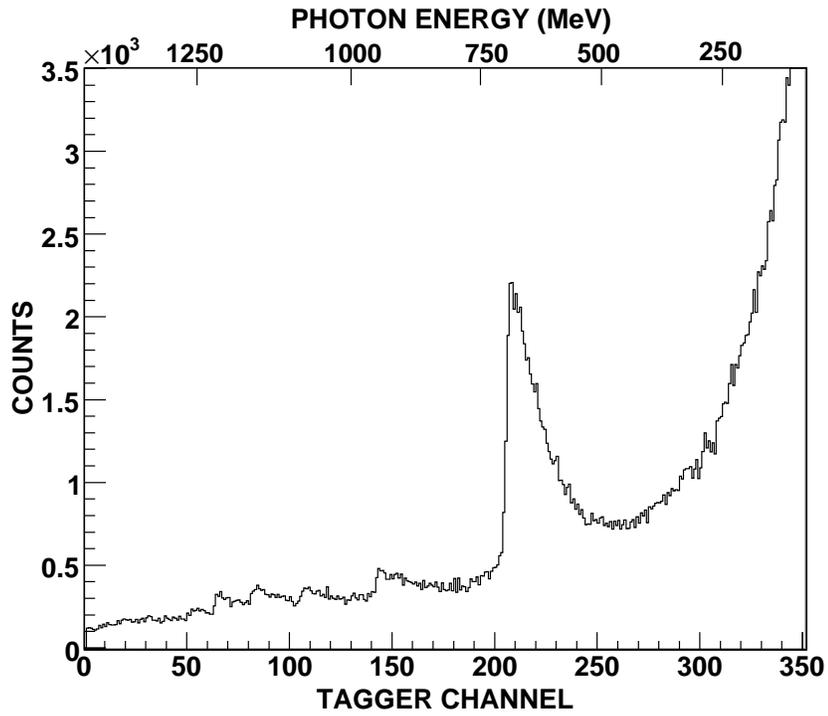}
\caption{Tagger spectrum in coincidence with a Pb glass detector placed
in the photon beam obtained using a 30 micron thick diamond radiator and 
a 1~mm diameter collimator.}
\label{fig:diamond}
\end{center}
\end{figure*}

Tests using a 30~micron thick diamond radiator and a 1~mm diameter
collimator have been done to check that the upgraded system can produce
linearly polarised photons.
The tagger spectrum observed in coincidence with a Pb glass detector
placed in the photon beam is shown in Fig. \ref{fig:diamond}. The
diamond angles were set so that the main coherent peak is at a photon
energy of 680~MeV. From the height of the coherent peak above the
incoherent background the degree of linear polarisation in the peak
channels is seen to be $\sim$65\% (after taking account of the fact
that the coherent radiation is not 100\% polarised by making use of
equations 108 in ref \cite{poln2}).

\section{SUMMARY}
\label{sec:sum}

The upgrade of the Glasgow photon tagging spectrometer at Mainz
has been completed successfully. When used with the 1508 MeV beam
from MAMI-C and the main focal plane detector it provides tagged photons in the
energy range 80-1401~MeV with a photon flux up to $\sim$2.5x10$^5$ photons
per MeV and energy resolution of $\sim$4~MeV.
Energy calibration has been made using the accurately known MAMI
energies. 

The upgraded spectrometer has been in regular use for tagged-photon
experiments with the Crystal Ball and TAPS since the beginning of 2007,
using incident energies of 883~MeV and 1508~MeV. Experiments to
investigate $\eta$ and K meson photoproduction on the proton have already
yielded very promising preliminary results and will continue with
various polarised and unpolarised targets.

\begin{acknowledgement}
We would like to thank R. Thomson, Department of Mechanical
Engineering, University of Glasgow for assistance in setting up the
stress analysis, S. Kruglov and the technicians at the
Petersburg Nuclear Physics Institute, St. Petersburg,
for cutting the new scintillators, V. Bekrenev (St. Petersburg) V. Lisin
and S. Cherepnya (Moscow) for testing the PMTs and
J. McGavigan (Glasgow University), R. Hoffman and K.S. Virdee
(George Washington University)
for laboratory assistance in testing the amplifier/discriminator cards.
We also thank D. Doak, Science Faculty workshop, Glasgow University
and the workshop staff in Mainz for technical assistance, C-H. Kaiser,
A. Jankoviak and the Mainz accelerator group for providing electron beam
of excellent quality at several energies and K. Livingston (Glasgow)
for providing Fig. \ref{fig:diamond}.
This research was supported by the UK EPSRC, the Deutsche
Forschungsgemeinschaft (SFB 443) and DAAD (ARC-program ARC-X-96/21)
and is part of the EU integrated infrastructure initiative hadron physics
project under contract number RII3-CT-2004-506078.
\end{acknowledgement}
 
%

%
%

%

%

\end{document}